# Stock Price Forecasting in Presence of Covid-19 Pandemic and Evaluating Performances of Machine Learning Models for Time-Series Forecasting

Navid Mottaghi
*Department of Physics and Astronomy, West Virginia University, Morgantown, West Virginia 26506, USA.*
namottaghi@mix.wvu.edu

Sara Farhangdoost
*Division of Resource Economics and Management, West Virginia University, Morgantown, West Virginia 26505, USA.*
sf0041@mix.wvu.edu

*Abstract:* With the heightened volatility in stock prices during the Covid-19 pandemic, the need for price forecasting has become more critical. We investigated the forecast performance of four models including Long-Short Term Memory, XGBoost, Autoregression, and Last Value on stock prices of Facebook, Amazon, Tesla, Google, and Apple in COVID-19 pandemic time to understand the accuracy and predictability of the models in this highly volatile time region. To train the models, the data of all stocks are split into train and test datasets. The test dataset starts from January 2020 to April 2021 which covers the COVID-19 pandemic period. The results show that the Autoregression and Last value models have higher accuracy in predicting the stock prices because of the strong correlation between the previous day and the next day's price value. Additionally, the results suggest that the machine learning models (Long-Short Term Memory and XGBoost) are not performing as well as Autoregression models when the market experiences high volatility.

I. INTRODUCTION

The non-linearity and unpredictability nature of the stock market make the decision hard on predicting the stock prices. A prediction is an important tool for investors to have successful trades which gives confidence to traders to know when to buy and sell the stocks in appropriate time frames. Forecasting stock prices is important to market participants since a reasonably accurate prediction can yield high financial benefits and to policymakers who analyze the market impact of domestic or international events.

The successful prediction can give desirable profit for investors, however, many factors such as pandemics, and social movements can have significant effects on market prices. The effects on the stock market have been observed recently due to the COVID-19 pandemic. The spread of the virus in the world had and still has huge economic impacts. These effects are evident which can be seen from the history of the volatility in the United States from 1929 to the present. The volatility rapidly increased to 0.05 which was more than other pandemics and economic depressions such as the Great Depression in 1933 and Global Financial Crisis in 2008 [1].

With the heightened volatility in stock prices especially during the Covid-19 pandemic, the need for price forecasting has become even more critical. Due to the high stock price fluctuations, many stock investors have lost all their property by wrong investments during the Covid-19 pandemic. On the other hand, accurate forecasting tools and sufficient market information help market participants in making both long- and short-term profits.

Several existing papers have analyzed the stock market prices. For instance, Tsai and Wang [2] found that the combination model of Artificial Neural Networks (ANN) and decision trees (DT) has better performance in predicting the stock price in the electronic industry in Taiwan than the single ANN and DT models. Similarly, Wu and Duan [3] examined the forecast performance of several ANN models and indicate that the Elman Neural Network model has an obvious advantage over the other models. Du [4] shows that for forecasting the Shanghai Securities Composition stock index is better to use nonlinear models such as ARIMA-BP Neural Network to improve the accuracy of prediction. The previous papers mostly focus on comparing the forecast performance of different models. However, none of these papers examine the accuracy of these forecasting models in a period that the market experience high volatility such as the Covid-19 pandemic period. Therefore, in the present paper, we investigate the forecast accuracy of different models during the Covid-19 pandemic period.

The purpose of this study is: First, to understand the COVID-19 pandemic effects on the stock market. Second, we evaluate how well different price forecasting models including the Autoregression model, Long Short-Term Memory (LSTM), XGBoost, and Last-Value can predict specific stock prices such as Facebook, Google, Microsoft, Apple, Tesla, and Amazon.

II. METHODOLOGY

A. *Autoregression Model*

The Autoregression methodology is a well-known model which is widely used in time series analysis. This model uses the data from previous time steps as input variables to the regression equation to predict the variable of the next time steps. For example, the variable at time t is $y_t$ which is an input to the model to predict the value of the next variable at time $y_{t+1}$ where the number of lags is 1. In this model it is possible to predict the value of $y_{t+1}$ from the last two-time steps, $y_{t-1}$ and $y_t$ meaning the number of lags is 2. In this model, the correlation between the variables is important since forecasting the future values are highly dependent on the past values [5]. The correlation depends ($r_k$) on the number of lags (k) and it can be defined as

$$r_k = \frac{\sum_{t=k+1}^{T}(y_t - \bar{y})(y_{t-k} - \bar{y})}{TS^2}, \qquad \text{(II-1)}$$

where T is the time period and $\bar{x}$ is defined as the sample mean

$$\bar{y} = \bar{T} \sum_{t=1}^{T} y_t, \quad \text{(II-2)}$$

and

$$s^2 = T^{-1} \sum_{t=1}^{T} (y_t - \bar{y})^2, \quad \text{(II-3)}$$

$s^2$ as the variance of $y_t$. The lag plots of all stocks are shown in Figure 1. The plots show a linear pattern with a positive slope suggesting positive autocorrelation between two consecutive times: $y_t$ and $y_{t+1}$. The red dashed diagonal line is plotted to present the linear trend of data.

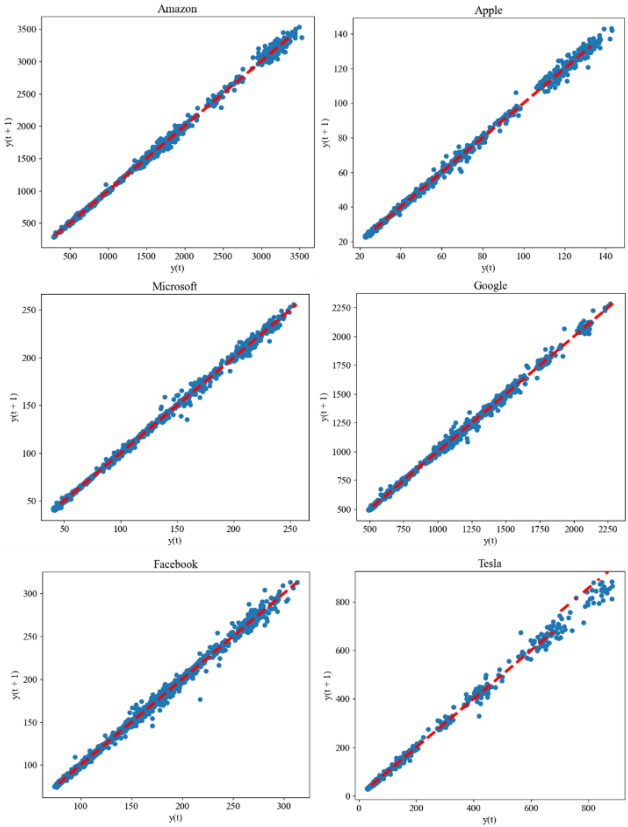

Figure 1. Lag plots of Amazon, Apple, Microsoft, Google, Facebook, and Tesla.

The strong positive correlation between $y_{t+1}$ and $y_t$ suggests the Autoregression model can be a good one to study the time variation of stock prices.

### B. LSTM

Long Short-Term Memory (LSTM) is a model based on a neural network that can be used in time series analysis. LSTM uses the past time variables as observations and it has multiple components including the memory cell, forget gate, and input gate. The memory cells are changeable based on the input and forget gates. The structure of the LSTM is suitable for gradients to move across many time-steps while the information stays in many time-steps. Each unit has two types of connections: the previous time-steps and previous layers [6].

### C. XGBoost

The eXtreme Gradient Boosting (XGBoost) is an implementation of gradient boosted decision trees. Gradient boosting is a technique that reduces the errors of existing models by adding new models and it uses a gradient descent algorithm to minimize the loss. The ease of implementation of the model with its high performance on difficult machine learning (ML) tasks makes it an excellent ML approach [7].

### D. Last Value

The last value or random walk model is based on the efficient market hypothesis which states that market prices should fully reflect all past and present available public and private information under the weak, semi-strong, and strong form of market efficiency, respectively [8]. There are several papers in the literature [6, 7] that show that previous-period price often performs well against more complicated forecasting models, and the last value model mostly used as the benchmark for forecast comparisons.

## III. DESCRIPTION OF THE DATA

The observation of the recent volatility in the USA stock market motivates us to study the different models to predict the stock price of famous companies such as Microsoft, Tesla, Facebook, Google, Apple, and Amazon. We use daily stock prices including the price from Yahoo Finance data source between Jan 2015 to April 2021. The data includes the High, Low, Open, Close, Adjacent close, and Volume which in this study only the closing price has been used.

To train the forecasting models, the price of each stock before January 2020 is used and the model's prediction ability is tested using data from January 2020 to April 2021. The statistics of the dataset are summarized in Table 1.

Table 1. Summary Statistics of Close Price

| | Mean | Min | Max | SD | No.Obs |
|---|---|---|---|---|---|
| **Microsoft** | | | | | |
| Training Set | 81.31 | 40.29 | 160.62 | 32.29 | 1262 |
| Test Set | 202.06 | 135.41 | 255.85 | 26.42 | 316 |
| **Apple** | | | | | |
| Training Set | 38.73 | 22.58 | 75.08 | 11.19 | 1262 |
| Test Set | 102.59 | 56.09 | 143.16 | 23.71 | 316 |
| **Tesla** | | | | | |
| Training Set | 53.89 | 28.73 | 93.81 | 11.39 | 1262 |
| Test Set | 389.33 | 72.24 | 833.09 | 241.76 | 316 |
| **Google** | | | | | |
| Training Set | 915.09 | 491.20 | 1394.21 | 232.13 | 1262 |
| Test Set | 1593.99 | 1056.62 | 2285.88 | 273.57 | 316 |
| **Amazon** | | | | | |
| Training Set | 1117.52 | 286.95 | 2039.51 | 535.89 | 1262 |
| Test Set | 2796.26 | 1676.61 | 3531.45 | 521.74 | 316 |
| **Facebook** | | | | | |
| Training Set | 143.28 | 74.05 | 217.5 | 38.21 | 1262 |
| Test Set | 242.95 | 146.01 | 313.09 | 38.39 | 316 |

For each stock, we report the mean value, maximum value, minimum value, and standard deviation (SD) for the training and test datasets, separately. The mean value of test datasets for all stocks is higher than the mean value of the training datasets. For example, the mean value of the training set and test sets for Microsoft, Tesla, and Amazon are $81.31, $53.98, $1117.52, and $202.06, $389.33, $2796.26, respectively. The higher mean value of the test set shows that the market had a jump from November 2020 to April 2021. The SD values of stocks show that the dispersion of price values is high indicating the values are spread out. Amazon, Tesla, and Google have the first, the second, and the third rank in the highest SD among the stocks. The price jump in the stock market in the test time frame implies the effect of the first COVID-19 wave in the market which increases the mean, minimum and maximum values of prices in the test set compared with the training set.

## IV. RESULTS AND DISCUSSIONS

The stock prices with the models within the time frame of January 2020 to April 2021 are shown in Figure 2. The provided plots are the test data to present the models' prediction on the stock prices. As it is discussed in the introduction the COVID-19 has a tremendous impact on the stock market thus for convenience the dates when the number of case peaks is indicated in the graph of Tesla. These dates are obtained Forbes where each Covid-19 wave had the maximum number of cases. The important note here is the effect of the virus in increasing the stock price which causes volatility in the market. Thus, the high volatile time frame is the interest of this study and the predicted time frame is chosen from April 2020 to April 2021. The best performance of models is by the Autoregression, Last Value, and XGBoost models which can be seen from Figure 2. On July 11, 2020, the LSTM model prediction has a deviation from the prices of all stock except Google, however, this model could not follow the trend from October 24, 2020, on Google price which implies that this model is not suitable in predicting stock prices for high volatile regions.

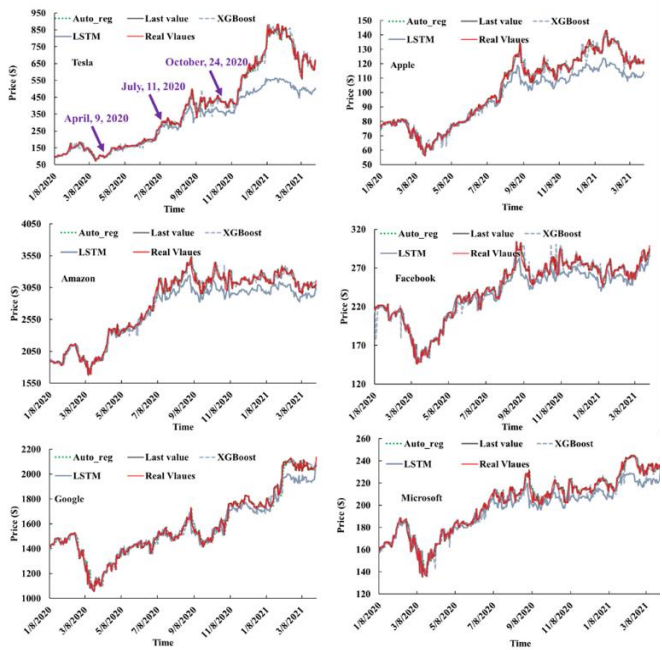

Figure 2. The plot of Competing Forecasts ($) between Jan. 2020 to Apr.2021.

To have better insight into the model performances, we listed the mean absolute error (MAE) root mean squared error (RMSE) values of all models in Table 2. Among the models, the Autoregression and Last-Value models work, the best which is expected since the price values of two consecutive days have a positive correlation (Figure 1). For example, the MAE and RMSE values of the Autoregression for Microsoft, Apple, and Facebook are 3.3239, 1.92, 4.5977, and 4.5878, 2.6160, 6.1932, respectively. The Last Value has also fewer MAE and RMSE values which close to the Autoregression model. For instance, these MAE values for the mentioned stocks are 3.3433,1.9204 4.0618 and, RMSE values are 4.6018, 2.6256. 6.2081. In general, the LSTM model has the worst performance. It is worth saying that the SD of prices is high that Amazon and Google have the highest values which makes the prediction hard for the models.

Table 2. Forecasting Error for Each Forecasting Method between Jan. 2020 to Apr. 2021 (Test Set)

|  | Forecasting Models | Auto-regression | LSTM | XGBoost | Last value |
|---|---|---|---|---|---|
| **MAE** | Microsoft | 3.3239 | 6.0038 | 4.192 | 3.3433 |
|  | Apple | 1.9200 | 6.1825 | 2.513 | 1.9204 |
|  | Tesla | 13.8713 | 75.5746 | 19.108 | 13.6959 |
|  | Google | 24.4496 | 36.9705 | 28.275 | 24.3832 |
|  | Amazon | 46.2729 | 119.7464 | 58.220 | 46.3224 |
|  | Facebook | 4.5977 | 7.7448 | 5.848 | 4.6018 |
| **RMSE** | Microsoft | 4.5878 | 8.0697 | 5.661 | 4.6081 |
|  | Apple | 2.6160 | 8.4841 | 3.438 | 2.6256 |
|  | Tesla | 20.6022 | 119.69 | 27.806 | 20.4481 |
|  | Google | 34.2613 | 53.5234 | 39.491 | 34.1388 |
|  | Amazon | 61.5987 | 157.3958 | 75.531 | 61.8686 |
|  | Facebook | 6.1932 | 10.3954 | 8.194 | 6.2081 |

## V. CONCLUDING REMARKS

From the modeling investigations on the selected stock prices, four major results have been concluded. First, the COVID-19 had a noticeable impact on the market that the price of the market increased after each wave of this virus reached its highest number of cases which resulted in high volatility. For example, after October 24, 2020, the price of Tesla, Apple, and Google significantly increased. Second, the statistical analysis of the stock prices was split into two major times one before the spread of COVID-19 and the second one after the time to understand the effect of this virus. From January 2020 to April 2021, after this virus started to spread, that the Mean and SD of price are substantially higher than before January 2020. Third, the Autoregression and the Last Value models have the best performance as both have the lowest MAE and RMSE values compared with the XGboost and LSTM models. Fourth, the strong correlation between the two successive days (Figure 1) yields a better prediction of Autoregression and Last Value models.